\documentclass[pdflatex,sn-mathphys-num]{sn-jnl}

\usepackage{graphicx}%
\usepackage{multirow}%
\usepackage{amsmath,amssymb,amsfonts}%
\usepackage{amsthm}%
\usepackage{mathrsfs}%
\usepackage[title]{appendix}%
\usepackage{xcolor}%
\usepackage{textcomp}%
\usepackage{manyfoot}%
\usepackage{booktabs}%
\usepackage{algorithm}%
\usepackage{algorithmicx}%
\usepackage{algpseudocode}%
\usepackage{listings}%
\usepackage{lineno}

\theoremstyle{thmstyleone}%
\theoremstyle{thmstyletwo}%

\theoremstyle{thmstylethree}%

\newcommand{\yambo} {{\normalfont\ttfamily Yambo}}
\begin{document}

\title[Article Title]{Ultrafast nonlinear Hall effect in black phosphorus}


\author*[1,2]{\fnm{Maciej} \sur{Dendzik}}\email{dendzik@kth.se}

\author[3]{\fnm{Andrea} \sur{Marini}}

\author[1,4]{\fnm{Samuel} \sur{Beaulieu}}

\author[1,5]{\fnm{Shuo} \sur{Dong}}

\author[1,6]{\fnm{Tommaso} \sur{Pincelli}}

\author[1]{\fnm{Julian} \sur{Maklar}}

\author[1]{\fnm{R. Patrick} \sur{Xian}}

\author[3,7]{\fnm{Enrico} \sur{Perfetto}}

\author[1]{\fnm{Martin} \sur{Wolf}}

\author[7,8]{\fnm{Gianluca} \sur{Stefanucci}}

\author[1,6]{\fnm{Ralph} \sur{Ernstorfer}}

\author*[1]{\fnm{Laurenz} \sur{Rettig}}\email{rettig@fhi-berlin.mpg.de}

\affil[1]{\orgname{Fritz Haber Institute of the Max Planck Society},
\orgaddress{\street{Faradayweg 4--6}, \city{Berlin}, \postcode{14195}, \country{Germany}}}

\affil[2]{\orgdiv{Department of Applied Physics},
\orgname{KTH Royal Institute of Technology},
\orgaddress{\street{Hannes Alfvéns väg 12}, \city{Stockholm}, \postcode{114 19}, \country{Sweden}}}

\affil[3]{\orgname{CNR-ISM, Division of Ultrafast Processes in Materials (FLASHit)},
\orgaddress{\street{Via Salaria Km 29.3}, \city{Monterotondo Scalo}, \postcode{00016}, \country{Italy}}}

\affil[4]{\orgname{CELIA, University of Bordeaux--CNRS--CEA},
\orgaddress{\city{Bordeaux}, \country{France}}}

\affil[5]{\orgname{Beijing National Laboratory for Condensed Matter Physics, Chinese Academy of Sciences},
\orgaddress{\city{Beijing}, \postcode{100190}, \country{China}}}

\affil[6]{\orgname{Institute for Optics and Atomic Physics, Technical University Berlin},
\orgaddress{\city{Berlin}, \country{Germany}}}

\affil[7]{\orgdiv{Department of Physics},
\orgname{Tor Vergata University of Rome},
\orgaddress{\street{Via della Ricerca Scientifica 1}, \city{Rome}, \postcode{00133}, \country{Italy}}}

\affil[8]{\orgname{INFN, Sezione di Roma Tor Vergata},
\orgaddress{\street{Via della Ricerca Scientifica 1}, \city{Rome}, \postcode{00133}, \country{Italy}}}


\abstract{The nonlinear Hall effect (NHE) is a recently discovered member of the Hall effect family in which the Hall voltage shows a nonlinear behavior when a transverse electric field is applied. While the NHE does not require broken time-reversal symmetry, such as that induced by a magnetic field, it requires broken inversion symmetry, which limits the range of suitable systems and potential applications. Here, we demonstrate an ultrafast NHE in centrosymmetric black phosphorus through dynamical symmetry breaking using femtosecond light pulses. We provide a detailed microscopic picture of excited carrier dynamics and induced fields using momentum-resolved photoemission spectroscopy combined with \textit{ab-initio} calculations. The ultrafast NHE is observed exclusively for the light polarization aligned with the armchair high-symmetry direction and persists over 300~fs, which opens new possibilities for selective and ultrafast light-to-current conversions.}

\maketitle


The Hall effect family constitutes a cornerstone of modern condensed matter physics, significantly contributing to our understanding of solid-state materials and their electronic and transport properties~\cite{Hall79,Klitzing80,Liu16,Nagaosa10}. In the most general terms, Hall effects can be described as symmetry-broken states; therefore, fundamental symmetries and ways to manipulate them are of paramount importance. For example, the Onsanger reciprocity relations dictate that a resistivity tensor of a material described by a time-reversal invariant Hamiltonian is strictly symmetric~\cite{Zeng20}. Therefore, a magnetic field has often been utilized to induce the transverse Hall response. Strategies for realizing Hall effects without breaking time-reversal symmetry include the use of additional quantum degrees of freedom, such as spin~\cite{Yang11} and valley~\cite{Xiao07,Mak14}, or by surpassing the linear response regime~\cite{Ma19}. 

The nonlinear Hall effect (NHE) is a recently discovered phenomenon characterized by a second-order response to an external electric field at both zero and twice the driving frequency~\cite{Sodemann15,Moore10}. Such a response has a quantum origin as it arises from the nonzero Berry curvature dipole moments in the reciprocal space. In addition to this intrinsic origin, effects induced by extrinsic disorder, such as side jump and skew-scattering mechanisms, can also lead to NHE~\cite{Du21a}. Regardless of origin, NHE can be used to rectify an oscillating electric field to a direct current (DC) without a semiconductor p-n junction~\cite{Isobe20}. This behavior can be applied to low-power energy harvesters or efficient terahertz/infrared photodetectors\cite{Du21,Zhang21}. Furthermore, the nonlinear spin Hall effect has recently been proposed as an efficient route towards the generation of spin currents~\cite{Hayami22}.      

Observation of NHE typically requires a system with sufficiently low symmetry. Although breaking the time-reversal symmetry is not required for the generation
of a second-order nonlinear response, broken inversion symmetry is still necessary for steady-state NHE to occur~\cite{Sodemann15}. This limitation can be
transiently circumvented by exciting centrosymmetric crystals with short light pulses. Generally, the light field breaks the symmetries of a quantum
system's Hamiltonian on the time scale of the laser pulse~\cite{Ivanov93}. The concept of coherent symmetry manipulation using femtosecond light pulses has
become an important topic in nonlinear optics, with key examples such as ultrafast symmetry switch in Weyl semimetals~\cite{Sie19}, or generation of petahertz
nonlinear currents in centrosymmetric organic superconductors~\cite{Kawakami20}. However, the theory behind NHE is often discussed for frequencies below the
interband transition threshold ($\omega<10^{14}$ Hz) due to simplified Drude-like modeling of rectified DC currents originating from the Berry curvature dipole~\cite{Zhang21}. For optical excitation frequencies, the description is much more complex and requires modeling a realistic electronic band structure and interband transitions involved. Furthermore, a typical experimental approach to observe NHE by measuring transverse currents can be obfuscated by several competing optical effects, such as the bulk photovoltaic effect~\cite{Dai23,Morimoto16}. Therefore, material-specific simulations of ultrafast carrier dynamics and novel experimental methodology are vital to developing a better understanding of nonlinear light-to-current conversions.       

Previous experimental insight into NHE comes mainly from transport measurements or terahertz spectroscopy~\cite{Ma19,Gao23,Min23,Hu23}. However, these cannot
directly measure the state-resolved electron dynamics underlying the NHE and other nonlinear transport effects, because the typical time scales of scattering
events are usually in the range of femtoseconds. Electronic equipment typically used for conductivity measurements cannot produce trigger signals or detect
transients on these timescales. Recent developments in time- and angle-resolved photoemission spectroscopy (trARPES) allow for investigations of intra- and
intervalley scattering, electron-phonon coupling, or Berry curvature, which are necessary to understand and simulate nonlinear effects in quantum
materials~\cite{Na19,Dong21,Beaulieu24}. 

In this work, we used the trARPES technique to study the excited state dynamics of centrosymmetric black phosphorus (BP) in momentum space. We observed a population imbalance within pairs of excited states with opposite momentum indicative of an unexpected generation of ultrafast traverse currents, which we interpret as a clear manifestation of the novel photo-induced NHE. Based on our {\it ab-initio} modeling, we attribute the observed effect to induced internal electric fields that redistribute carriers depending on their crystal momentum. Interestingly, the induced transverse currents persist even after optical excitation, lasting up to $\sim$300~fs, marking the time when the symmetry of the system is restored.

BP~\cite{Ling15} has been identified as an auspicious candidate for optoelectronic applications due to its direct thickness-dependent bandgap (0.3-2~eV), high carrier mobility and compatibility with existing methods of large-scale production~\cite{Wu21}. The wide scientific interest in studying BP has been mainly fueled by its peculiar physical properties, such as anisotropic Dirac states, bandgap renormalization due to alkali metal deposition~\cite{Kim15,Kim17,Kim17a,Ehlen18}, or sizeable bipolar pseudospin polarization~\cite{Jung20}. The strongly anisotropic crystal structure of BP leads to broadband optical absorption dependent on polarization~\cite{Yuan15,Li18,Chang19} and tunable electro-optical light polarization conversion~\cite{Biswas21}. Previous trARPES investigations of BP~\cite{Roth19,Chen19,Kremer21,Zhou23} as well as theoretical calculations~\cite{Low15,Yar23} have focused exclusively on the global conduction band minimum located around the center of the surface Brillouin zone ($\Gamma$ valley). Here, we performed trARPES investigations using a momentum microscopy approach~\cite{Medjanik17,Maklar20,Beaulieu21} in order to capture ultrafast electron dynamics throughout the whole surface Brillouin zone of BP. The results reveal a transient carrier population in previously overlooked side valleys, showing a distinct valley asymmetry of excited states having opposite momentum driven by linearly polarized light. Furthermore, the experimental results are complemented with a state-of-the-art simulation scheme based on the {\it ab-initio} Non-Equilibrium Green's Function Theory\,({\it Ai}-NEGF). The {\it Ai}-NEGF method is based on a solid merging of {\it ab-initio} Density Functional Theory (DFT) and  Non-Equilibrium Green's Function Theory~\cite{Stefanucci25,Marini2013} implemented in the \yambo\, code~\cite{AndreaMarini2009,Sangalli_2019}. More information about the experimental setup and theoretical calculations can be found in the Methods section.

\begin{figure}
\centering
	\includegraphics[width=0.66\textwidth]{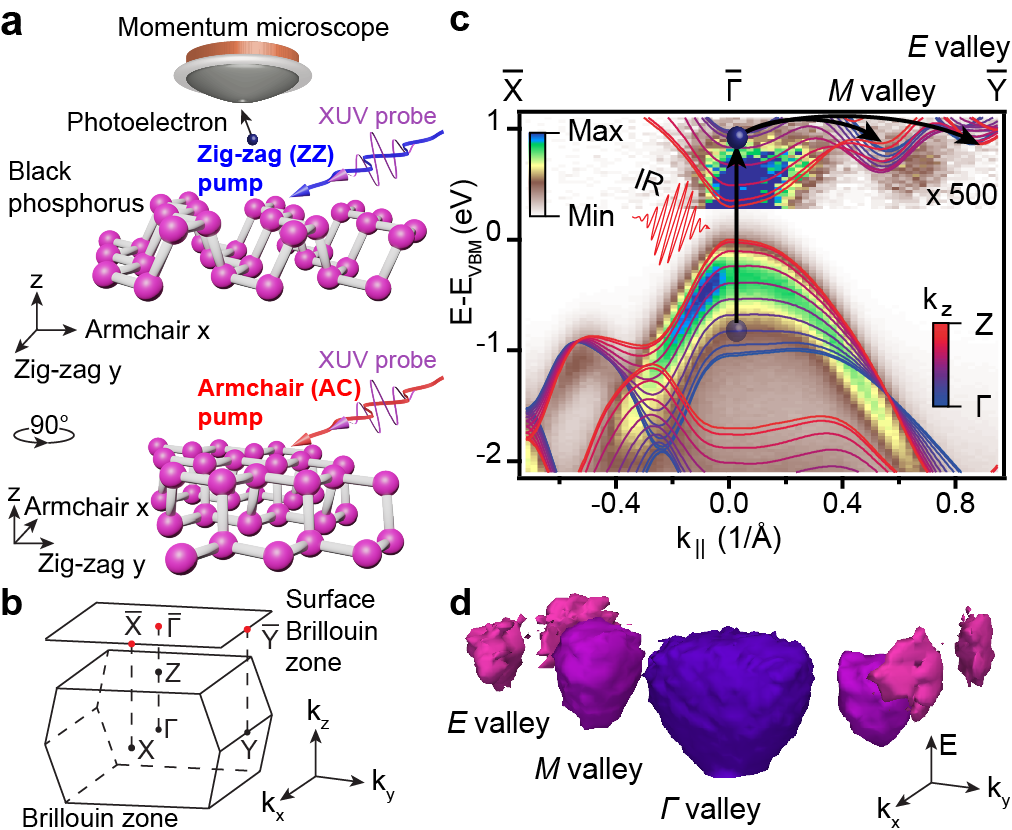}\\
	\caption{\textbf{Excited states mapping using time-resolved photoelectron momentum microscopy.}~\textbf{a}, A schematic of the experimental geometry including infrared pump pulses with linear polarization aligned with armchair (AC) and zig-zag (ZZ) directions and extreme ultraviolet (XUV) probe pulses.~\textbf{b}, Brillouin zone and projected surface Brillouin zone of BP.~\textbf{c}, trARPES spectra along high-symmetry directions obtained at $\Delta t=$75~fs with superimposed $k_z$-resolved DFT calculations of the electronic bandstructure. The signal in the conduction band (CB) was multiplied by a factor of 500 for clarity. Straight and curved arrows represent direct optical excitation of the $\Gamma$ valley and intervalley scattering processes, respectively.~\textbf{d}, Isometric projection of 3D photoemission intensity $I(E,k_x,k_y)$ showing the central valley $\Gamma$ and side $M$ and $E$ valleys.}
	\label{fig:fig1}
\end{figure}


We start the discussion of the results with a schematic of the experimental setup presented in Fig.~\ref{fig:fig1}a. In brief, we used infrared pulses with energy of 1.55~eV (pulse duration ca. 35~fs FWHM, incident fluence of $\sim0.07$~mJ/cm$^2$) and a monochromatized high-order harmonic generation (HHG)–based extreme ultraviolet source operating at a repetition rate of 500~kHz and centered around 21.7~eV (20~fs FWHM) as a pump and probe pulses, respectively. The pump was linearly polarized either along the armchair (AC) or zig-zag (ZZ) crystallographic directions of BP, experimentally realized by rotating the samples by 90$^{\circ}$ along the $z$ axis (Fig.~\ref{fig:fig1}a). Photoemission spectra were acquired using a momentum microscope that provides photoelectron signals across the entire surface Brillouin zone (Fig.~\ref{fig:fig1}b) without the need of changing the experimental geometry and thus preserving photoemission matrix elements. Fig.~\ref{fig:fig1}c presents photoemission spectra along the main high-symmetry directions acquired at a pump-probe delay $\Delta t$ of 75~fs superimposed with the equilibrium band structure calculated using DFT. In addition to the global CB minimum situated at $\Gamma$, we observed the excited carrier population in the vicinity of two additional side valleys. These excited state populations result from intervalley scattering processes as only states in the $\Gamma$ valley can be directly excited (Fig.~\ref{fig:fig1}c). The momentum distribution and relative energies of the central $\Gamma$ valley, middle $M$ and edge $E$ valleys are presented in the form of an isometric projection of the photoemission intensity in Fig.~\ref{fig:fig1}d.  

To investigate valley-resolved dynamics, we integrated the transient photoemission intensity over regions of interest (ROIs) in $k$-space corresponding to the $\Gamma$, $M$, and $E$ valleys, as presented in Fig.~\ref{fig:fig2}a (exact definition of ROIs can be found in the Supplementary Material). We compared the experimental results with the transient $k$-resolved electronic occupation obtained with our {\it Ai}-NEGF calculation integrated over the same ROIs (Fig.~\ref{fig:fig2}b). The same procedure was carried out for both AC and ZZ excitation conditions, and the results obtained are presented in Fig.~\ref{fig:fig2}d-f. Overall, an excellent agreement between the photoemission experiment and parameter-free realistic calculations provides a thorough insight into ultrafast carrier dynamics in BP upon optical excitation. We observed rapid depopulation of the $E$ valley (ca. 0.3~ps, Fig.~\ref{fig:fig2}f) and a somewhat slower decay of $M$ (ca. 1~ps, Fig.~\ref{fig:fig2}e). During this process, excited carriers from the side valleys scatter back to the global CB minimum, as evidenced by the increasing signal observed within the $\Gamma$ valley (Fig.~\ref{fig:fig2}d). Interestingly, the observed dynamics differ substantially for the AC and ZZ excitations, especially for the $\Gamma$ valley. The relative excited carrier density is found to be much smaller for ZZ compared to the AC excitation geometry, in agreement with the previously observed anisotropy of the photoabsorption cross section~\cite{Yuan15}. However, differences in absorbed fluence alone cannot explain the observed disparity of excited carrier dynamics. Fig.~\ref{fig:fig2}c presents the initial distribution of excited carriers for AC and ZZ excitations obtained by artificially "turning off" electron-phonon coupling in the {\it Ai}-NEGF calculations, allowing us to visualize the momentum distribution of the optical absorption. The results indicate that the center and edges of the $\Gamma$ valley are preferentially pumped with AC and ZZ light polarizations, respectively. We conclude that this disparity of the initial distribution of excited states, together with the strongly anisotropic phonon dispersion and electron-phonon coupling in BP~\cite{Ling16}, affects the complex intervalley scattering pathways and leads to the observed disparity of the valley-resolved carrier dynamics.

\begin{figure}
	\includegraphics[width=1\textwidth]{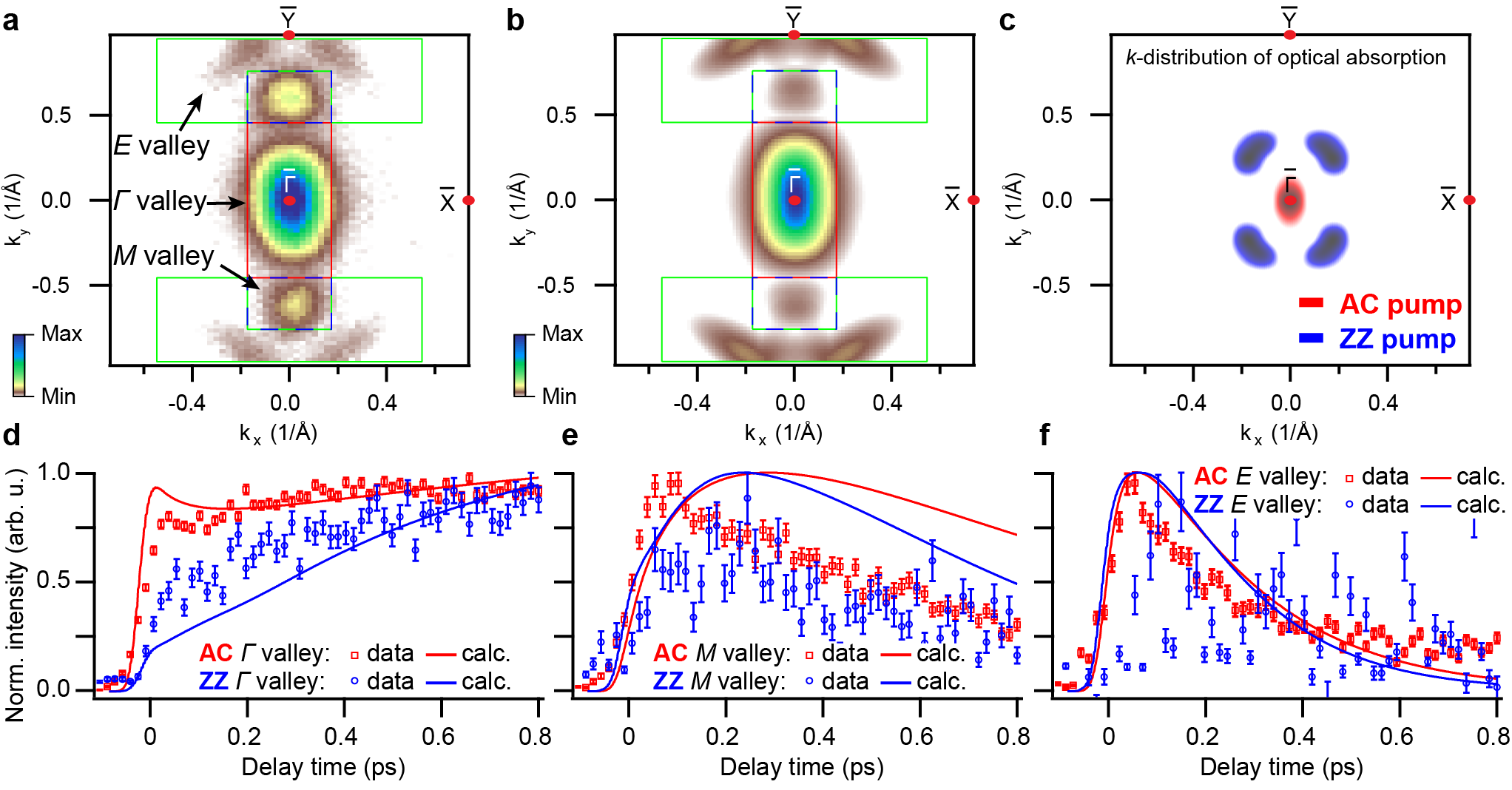}
	\caption{\textbf{Valley-resolved ultrafast electron dynamics.}~\textbf{a}, Energy-integrated (0.15-1.55~eV) momentum maps showing a 2D projection of the excited carriers population. Red, blue, and green lines mark the chosen region of interest (ROI) boundaries for the $\Gamma$, $M$ and $E$ valleys, respectively.~\textbf{b}, Calculated $k$-resolved occupation of the CB corresponding to experimental data presented in~\textbf{a}.~\textbf{c}, Calculated initial momentum distribution of excited states population using AC and ZZ pump without electron-phonon scattering.~\textbf{d-f}, Time-resolved photoemission spectra showing the observed populations in the $\Gamma$, $M$, and $E$ valleys for the AC (red) and ZZ (blue) pumping, respectively. Solid lines correspond to the calculated valley occupancy.} 
\label{fig:fig2}
\end{figure}

\begin{figure}
	\includegraphics[width=1\textwidth]{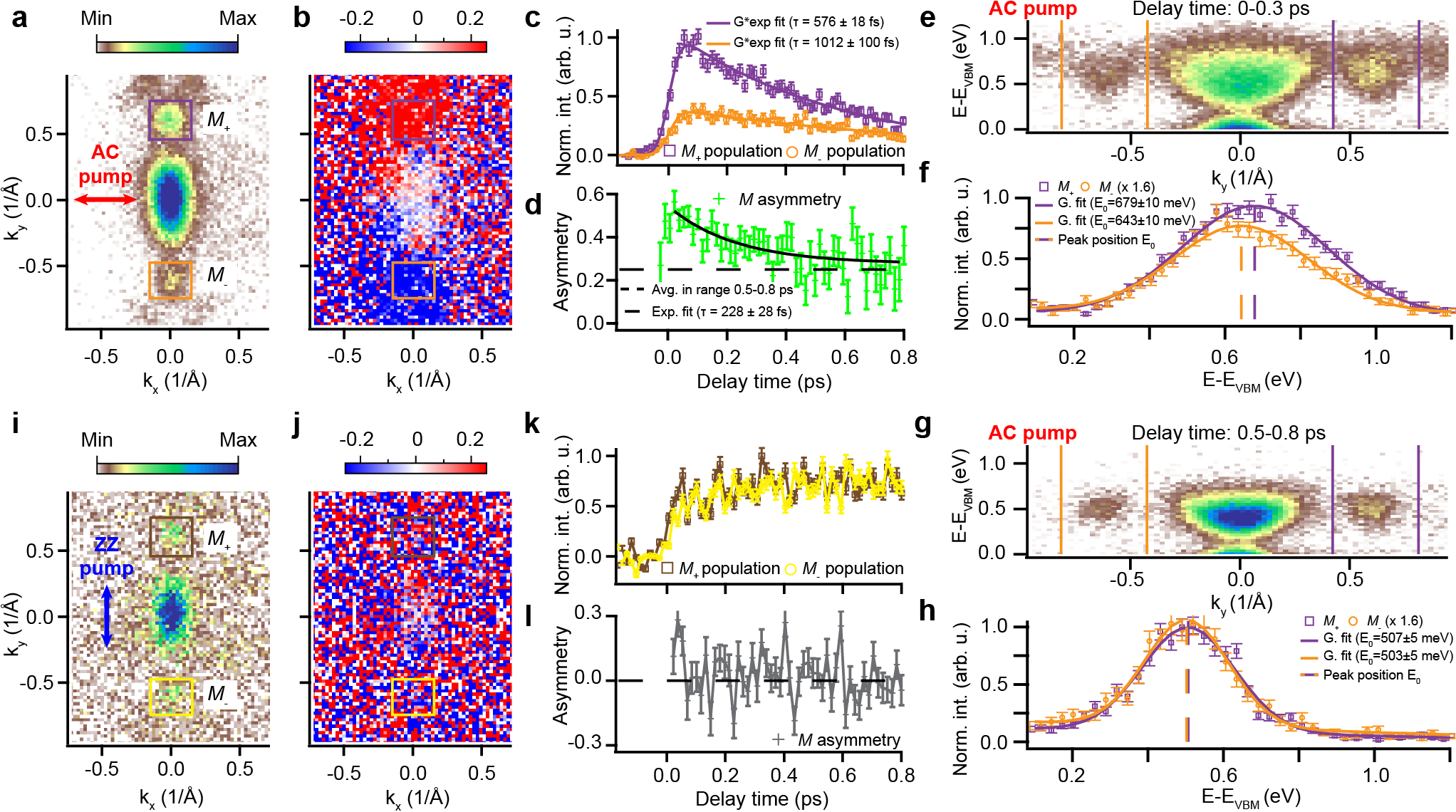}
	\caption{\textbf{Microscopic observation of the transient NHE in BP.}~\textbf{a}, Integrated (0.15-1.55~eV) constant energy map of excited states during the initial stage of the dynamics (0-0.3~ps) with indicated positions of $M_+$ (violet rectangle) and $M_-$ (orange rectangle) valleys measured with AC pump. The red arrow represents the AC direction equivalent to the $k_x$ direction.~\textbf{b}, Map of the asymmetry $\Delta I$ with indicated $M_+$ and $M_-$ valleys.~\textbf{c}, Dynamics of the $M_+$ and $M_-$ valleys together with exponential fits (see text). The fitting results are indicated in the insets.~\textbf{d}, Temporal evolution of the $\Delta I(M)$ asymmetry. The solid line marks an exponential fit, and the dashed line represents the average asymmetry at late times.~\textbf{e} and \textbf{g}, Photoemission spectra of the CB obtained with AC pumping for early ($\Delta t \in [0-0.3]$ ps) and late ($\Delta t \in [0.5-0.8]$ ps) delay times, respectively. Violet and orange lines indicate the positions of the $M_+$ and $M_-$ valleys, respectively. ~\textbf{f} and \textbf{h}, Valley-resolved energy distribution curves for $\Delta t \in [0-0.3]$ ps and $\Delta t \in [0.5-0.8]$ ps integrated in regions marked in \textbf{e} and \textbf{g}, respectively. Violet (orange) data points correspond to $M_+$ ($M_-$), and the solid lines mark the corresponding Gaussian fits. Dashed lines are placed at the fitted peak positions. $I(M_-)$ signal has been multiplied by a factor of 1.6 (see text).~\textbf{i-l}, Analogical graphs to \textbf{a-d} for the ZZ excitation conditions.} 
\label{fig:fig3}
\end{figure}

In addition to the distinct valley dependence, the observed carrier dynamics also exhibit a surprising dependence on the sign of the crystal momentum $k_y$ (Fig.~\ref{fig:fig3}). To quantify this effect, we introduce the zig-zag asymmetry factor $\Delta I(k_x,k_y) = \frac{I(k_x,k_{y})-I(k_x,-k_{y})}{I(k_x,k_{y})+I(k_x,-k_{y})}$ obtained from the photoemission intensity $I(k_x,k_y)$. Fig.~\ref{fig:fig3}b presents a map of $\Delta I(k_x,k_y)$ during the initial stage of dynamics (0-0.3~ps) obtained using AC pump. Although no significant asymmetry is observed for the $\Gamma$ valley, the side valleys show a distinct population imbalance. Due to rapid depopulation of the $E$ valley, we restrict the discussion to the $M$ valley and label states with $k_y>0$ and $k_y<0$ as $M_+$ and $M_-$, respectively (see Fig.~\ref{fig:fig3}a). Remarkably, the observed excited carrier populations $I(M_+)$ and $I(M_-)$ exhibit substantially different dynamics, as shown in Fig.~\ref{fig:fig3}c. A fit of an exponential decay convolved with the Gaussian instrumental response function reveals a significant difference in the decay constant $\tau$ of 576$\pm$18~fs and 1012$\pm$100~fs for $I(M_+)$ and $I(M_-)$. This results in a time-dependent asymmetry $\Delta I(M)$ as shown in Fig.~\ref{fig:fig3}d. The initial asymmetry of $\sim60\,\%$ decays on a time scale of $\sim200$~fs and settles at a final value of $\sim23\,\%$. The residual asymmetry observed for longer time delays can be ascribed to the forward-backward asymmetry of photoemission with respect to the photon axis originating from the probe propagation direction aligned with the $M$ valleys (the orbital term of photoemission matrix elements as defined in Ref.~\cite{Moser17}). However, a similar argumentation cannot explain the dynamics of the $M$ valley asymmetry because the time dependence of the photoemission matrix elements would require a dynamic change of the valley orbital character. Furthermore, while photoemission matrix elements can influence the intensity of the observed signal, it cannot change the energy distribution of excited carriers. We investigated the transient energetics of $M$-valley electrons obtained using AC excitation in Figs.~\ref{fig:fig3}e-h. Figs.~\ref{fig:fig3}e-f and Figs.~\ref{fig:fig3}g-h present $I(E,k_y)$ integrated during the initial (0-0.3~ps) and later (0.5-0.8~ps) delay times, respectively. The relative positions of the quasi Fermi-levels present in $M_+$ and $M_-$ are compared by analyzing the valley-resolved energy distribution curves (EDCs), as shown in Fig.~\ref{fig:fig3}f and Fig.~\ref{fig:fig3}h. We compensated for the experimental sensitivity difference arising from the photemission matrix elements by multiplying the EDC $M_-$ by a factor of 1.6, which corresponds to the long delay-time value of $\Delta I(M)$ (23\%), and fit the EDCs with phenomenological Gaussian distributions. The results reveal a significant difference in the observed central energy of $\Delta E_0=36\pm14$~meV (Fig.~\ref{fig:fig3}f) for the initial delay times. However, the same analysis performed for later delay times (0.5-0.8~ps) shows no difference in $\Delta E_0$, within the experimental uncertainty (Fig.~\ref{fig:fig3}h). 

Based on the above argumentation, we can safely exclude the effect of photemission matrix elements as the source of the experimentally observed carrier population imbalance. Note, that previous trARPES measurements showing an imbalance of excited states with opposite momentum have been attributed to the generation of ultrafast currents on the surfaces of metals~\cite{Gudde07} or topological insulators~\cite{Reimann18}. Therefore, we conclude that the observed $\Delta I(M)$ at short time delays corresponds to the generation of transient currents driven by side-valley asymmetry. Interestingly, the analogous measurements performed with pumping along the ZZ direction did not show observable electric current signatures (Fig.~\ref{fig:fig3}i-l). In this case, also the residual $\Delta I(M)$ is zero because the probe light propagation direction is perpendicular to the $\Gamma-M$ direction, resulting in the lack of the backward-forward photoemission asymmetry for the observed states.

\begin{figure}
	\includegraphics[width=1\textwidth]{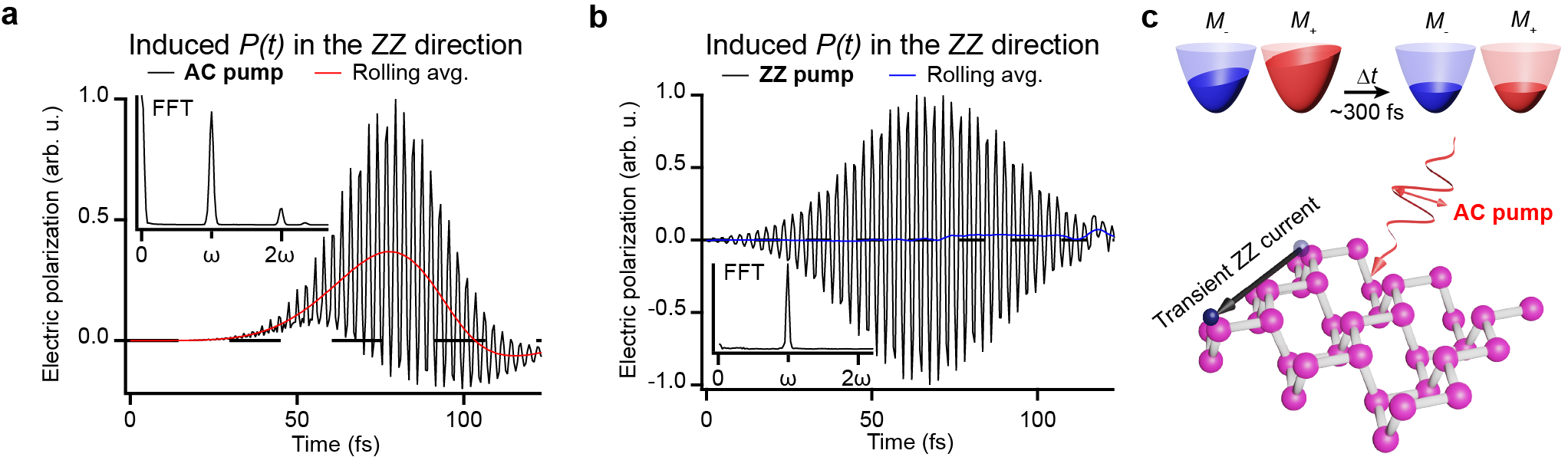}
	\caption{\textbf{Induced transient dielectric polarization inside of BP.}~\textbf{a}, Calculated induced polarization $P_y(t)$ along the ZZ direction for AC pumping. The red line marks the temporal average. The inset presents Fourier-transformed induced polarization with 0, $\omega$ and $2\omega$ components.~\textbf{b}, Calculated induced polarization $P_y(t)$ along the ZZ direction for ZZ pumping. The blue line marks the temporal average. The inset presents Fourier-transformed induced polarization with only the $\omega$ component.~\textbf{c}, A schematic illustration of the observed ultrafast NHE: AC pump generates the $M$ valley asymmetry that is manifested by a transient current along the ZZ direction.}   
\label{fig:fig4}
\end{figure}

To elucidate the origin of the observed ultrafast currents, we turn to our theoretical calculations. External laser-pulse excitation is known to simultaneously induce carrier scattering processes and time-dependent macroscopic electric polarization, $\mathbf{P}(t)$. In Fig.~\ref{fig:fig2} we have shown that the simulated carrier dynamics reproduces the experimental observations with high fidelity, and now we focus on the behavior of the light-induced polarization. Because BP is an anisotropic material, its dielectric response is tensorial. Consequently, regardless of the excitation geometry, the induced polarization generally does not need to align with the driving field. Using the present {\it Ai}-NEGF framework, we calculated $\mathbf{P}(t)$ for the two excitation configurations (AC and ZZ) matching the experimental conditions. The results are summarized in Fig.~\ref{fig:fig4}. When the laser field is aligned with the AC direction, we find a finite $P_y(t)$ along the ZZ direction (Fig.~\ref{fig:fig4}a). More specifically, the calculated $P_y(t)$ exhibits rapid oscillations at the optical frequency, superimposed on a slowly varying envelope that generates a net polarization within the first 100~fs. The Fourier spectrum of $P_y(t)$ reveals components at zero frequency and at the second harmonic of the driving field (inset of Fig.~\ref{fig:fig4}a), a characteristic signature of the NHE. The zero-frequency component transiently breaks the $k_y>0$ / $k_y<0$ population symmetry and corresponds to an ultrafast polarization of the side valleys (top panel of Fig.~\ref{fig:fig4}c). This directly accounts for the emergence of a transverse current and voltage perpendicular to the applied field, which are the defining hallmarks of the Hall effect (bottom panel of Fig.~\ref{fig:fig4}c). In addition, we find that the magnitude of the NHE is strongly polarization-dependent. When the pump is polarized along the ZZ direction, the induced polarization simply follows the oscillating driving field and averages to zero over an optical cycle (Fig.~\ref{fig:fig4}b). In this case, no valley asymmetry is generated, in full agreement with our experimental findings (Fig.~\ref{fig:fig3}i-l).

It should be mentioned that the valley-resolved carrier population simulations presented in Fig.~\ref{fig:fig2} show no $k_y>0$ and $k_y<0$ carrier population asymmetry. These calculations take into account only the external fields originating from optical excitation and not the induced polarization within the material. Including both external and induced polarizations in realistic simulations of carrier population in a fully self-consistent manner is beyond the scope of this study and will be a subject of separate investigations. However, the calculations of electric polatization $\mathbf{P}(t)$ presented here agree well with the experiment and serve as an explanation of the observed nonlinear effects. Furthermore, electric polarization has been shown to be equivalent to the Berry connection~\cite{King-Smith93,Resta94}. Therefore, we cautiously attribute the quantum geometry to be the main source underlying the observed NHE. As the Berry curvature vanishes under time-reversal and spatial-inversion symmetries, the induced polarization in centrosymmetric BP can be nonzero only transiently, a behavior that is well reproduced by both our calculations and experiments. Finally, it is also important to discuss the sign of the observed NHE voltage corresponding to the generation of a positive carrier population asymmetry for AC pumping. Taking into account the experimental conditions and symmetries present in our calculations, we conclude that the light propagation direction defines the sign of the zero-frequency component of the induced electric polatization and thus also the sign of the NHE voltage. Experimentally changing the polarity needs counter-propagating pump and probe pulses, which would require large modifications to the setup beyond the scope of the current work.   

In conclusion, our time- and momentum-resolved photoemission spectroscopy measurements enable a direct observation of electronic valley asymmetry equivalent to the conceptually new ultrafast nonlinear Hall effect. Our valley-resolved measurements carried out with two different light polarization directions (AC and ZZ) enable detailed benchmarking of first-principles microscopic calculations and, as a result, accurate calculation of macroscopic quantities such as electric polarization. Both theory and experiment indicate that the maximum NHE is obtained for AC pump polarization, which is explained in terms of initial population differences in the momentum distribution of excited carriers. We observed the ultrafast NHE by tracking the transient population and the energy distribution of electrons in the side valley $M$. In addition, the calculated induced polarization $P_y(t)$ shows Fourier components at zero and the second harmonic of the driving field -- a typical frequency response predicted for the NHE. Breaking of the inversion symmetry required for NHE is achieved transiently using femtosecond light pulses, and the sign of the Hall voltage depends on the light propagation direction. The work presented here provides the first microscopic obervation of ultrafast carrier dynamics accompanying the NHE and paves the way for future applications of black phosphorus in petahertz opto-electonics or light polarization detectors.

\section*{Methods}
\subsection*{Time-resolved momentum microscopy measurements}
The experimental setup consists of a light source based on high harmonic generation (HHG) in an argon jet delivering $\sim35$~fs pulses with an energy of 21.7~eV at a repetition rate of 500~kHz~\cite{Puppin19}. The probe light was \textit{p}-polarized for all measurements presented and nearly collinear with the pump beam (1.55~eV, 30~fs). The incoming excitation fluence was $\sim0.07$~mJ/cm$^2$ and the pump polarization was controlled using a half-wave plate. 

Photoemission spectra were acquired using a momentum microscope~\cite{Medjanik17} (METIS 1000, SPECS) with estimated energy resolution of ca. 150 meV. Before each measurement, a fresh surface of black phosphorus (BP) was prepared by \textit{in-situ} cleaving in ultra-high vacuum (UHV) with a base pressure of $<5\times10^{-11}$~mbar. The measurements were conducted at room temperature. The orientation of the samples was set by a 6-axis manipulator (Carving, SPECS) and monitored by observing constant-energy photoemission images. We estimate the uncertainty of the direction of pump light propagation with respect to the crystal axes to be $<2^\circ$. 

The recorded single-event data was binned into 4D datasets $I(k_x,k_y,E,t)$, using an open-source end-to-end workflow~\cite{Xian20}. Energy calibration was performed after every measurement by applying an additional reference voltage to the sample. Momentum calibration was performed after processing by fitting the observed electronic dispersion in the neighboring surface Brillouin zones and using reference values for the BP crystal structure (a=4.376~\AA, b=3.314~\AA)~\cite{Brown65}. Experimental artifacts, which led to minor deformations in the obtained photoemission spectra, were corrected as previously described~\cite{Xian19}.

\subsection*{First-principles calculations}

The first--principles simulation of the out--of--equilbrium carrier dynamics has been peformed with the \yambo\, code~\cite{AndreaMarini2009,Sangalli_2019}.
\yambo\, implements the Kadanoff--Baym Equation\,(KBE) of motion for the density matrix  $\rho_{nm\mathbf{k}}\left(t \right)$ written in the bands ($n,m$) and
$\mathbf{k}$ basis.

The KBE is obtained within the Non--Equilibrium Green's function approach~\cite{Stefanucci25} implemented, in \yambo, within the  Generalized Kadanoff Baym
ansatz\,(GKBA)~\cite{Marini2013,Sangalli2015a,Marini2021}. The GKBA allows to write the KBE equation of motion just in terms of the one time density
matrix  making possible to perform calculations in realistic materials composed of hundreds of atoms.

A typical \yambo\, calculations starts with the
Quantum--Espresso suite~\cite{Giannozzi_2009}, that is used to calculate the equilibrium properties of BP. In practice we have first performed a Density--Functional--Theory
calculation of the structural and electronic properties using the Perdew-Burke-Ernzenhof\,(PBE) approximation for the exchange--correlation functional. The BP
eigenvalues and wave--functions are calculated by using a 90~Ry  plane-waves energy cutoff and a $6\times 3\times 6$ $\bf{k}$--point mesh for the Brillouin zone
sampling. We then consider 4 valence and 4 conduction bands to describe the carrier dynamics.  Electron--phonon  matrix elements are computed within the
Density-Functional-Perturbation-Theory.  The density matrix is then
propagated in time by using, as external perturbation,  
a narrow--band pump laser pulse characterized by the experimental frequency, intensity, and width.

The real-time simulation
adopts a coarse $4\times 4\times 4$ and a denser randomly generated grid of 5000 $\bf{k}$--points. The actual simulation is performed on the denser grid with
matrix elements and density matrix interpolated (by using a nearest--neighbor technique) from the coarse grid.  
The screened--exchange\,(SEX) approximation guarantees that the
pump is correctly absorbed~\cite{Attaccalite2011}. Finally, we monitor the diagonal elements of the density matrix that describe the occupations in the band
structure, ${f_{n\mathbf{k}}=\rho_{nm\mathbf{k}}}$.

\bmhead{Supplementary information}
\bigskip\noindent
The online version contains supplementary material available at
\bmhead{Acknowledgements}
\bigskip\noindent
This work was funded by the Max Planck Society, the European Research Council (ERC) under the European Union’s Horizon 2020 research and
innovation program (Grant No. ERC-2015-CoG-682843), the German Research Foundation (DFG) within the Emmy Noether program (Grant No. RE 3977/1), through
Projektnummer 18208777-SFB 951 “Hybrid Inorganic/Organic Systems for Opto-Electronics (HIOS)” (CRC 951 project B12, M.S., D.C., A.K.), and the SFB/TRR 227
“Ultrafast Spin Dynamics” (projects B07, project-ID: 328545488), and the Program DFG SPP2244 (project-ID: 443366970). Program.  A.M acknowledges the funding
received from the European Union projects: MaX {\em Materials design at the eXascale} H2020-INFRAEDI-2018-2020/H2020-INFRAEDI-2018-1, Grant agreement n. 824143; {\em Nanoscience Foundries and Fine Analysis -- Europe | PILOT}  H2020-INFRAIA-03-2020, Grant agreement n. 101007417; {\em PRIN: Progetti di Ricerca di rilevante interesse Nazionale} Bando 2020, Prot. 2020JZ5N9M. S.B. acknowledges financial support from the NSERC-Banting Postdoctoral Fellowships and support from ERC Starting Grant ERC-2022-STG No.101076639. M.D. acknowledges financial support from the Göran Gustafsson Foundation, Swedish Research Council under Grant No: 2022-03813 and the Carl Trygger Foundation. G.S. and E.P acknowledge funding from Ministero Università e Ricerca PRIN under grant agreement No. 2022WZ8LME, from INFN through project TIME2QUEST, from European Research Council MSCA-ITN TIMES under grant agreement 101118915, and from Tor Vergata University through project TESLA. Funded by the European Union. Views and opinions expressed are however those of the author(s) only and do not necessarily reflect those of the European Union. Neither the European Union nor the granting authority can be held responsible for them.
\bmhead{Author contribution}
\bigskip\noindent
M.D., S.B., S.D., T.P., J.M., and L.R. performed the trARPES measurement. M.D. analyzed the data and wrote the first draft of the
manuscript. R.E., L.R., and M.W. were responsible for developing all the experimental infrastructures. A.M., E.P. and G.S. performed the theoretical
calculations. R.P.X. and developed the 4D data processing code. M.D., L.R. and R.E. conceived the project. All authors contributed to the final version of the
manuscript.
\bmhead{Conflict of interest}
\bigskip\noindent
The authors declare no competing interests.

\bmhead{Data availability}
\bigskip\noindent
All experimental data and calculations shown in the main text are accessible at Zenodo.

\end{document}